\documentclass[a4paper,11pt]{article}
\pdfoutput=1 
\usepackage{jheppub} 
 \usepackage{graphicx}
 \usepackage{subfig}
 \usepackage{caption}
 \usepackage{subcaption} 
\graphicspath{ {./} }                    
\usepackage{slashed}
\usepackage[T1]{fontenc}
\usepackage{amssymb}
\usepackage{mathrsfs}
\usepackage{amsmath}
\usepackage{dsfont}
\usepackage{float}
\usepackage{hyperref}
\usepackage{cancel}
\usepackage[normalem]{ulem}
\usepackage{xcolor}
\usepackage{color}
\usepackage{braket}
\usepackage{bm}
\usepackage{calligra}
\usepackage{feynmp}
\usepackage{tikz}
\usepackage{tikz-feynman}
\usepackage{orcidlink}
\tikzfeynmanset{compat=1.1.0}
\usetikzlibrary{positioning}
\newcommand{\beq}{\begin{equation}}
\newcommand{\eeq}{\end{equation}}
\def\barr{\begin{array}}
\def\earr{\end{array}}

\newcommand{\bsp}{\begin{split}}
\newcommand{\esp}{\end{split}}
\newcommand{\bit}{\begin{itemize}}
\newcommand{\eit}{\end{itemize}}

\definecolor{darkcyan}{cmyk}{1,0,0,0.4}
\definecolor{darkgreen}{cmyk}{1,0,1,0.4}

\def\lapp{\mathrel{\rlap{\raise.5ex\hbox{$<$}}
                    {\lower.5ex\hbox{$\sim$}}}}
\def\gapp{\mathrel{\rlap{\raise.5ex\hbox{$>$}}
                    {\lower.5ex\hbox{$\sim$}}}}
\begin{document}

\title{\boldmath A relook at a low reheating temperature: Freeze-in versus Freeze-out}

\author[a]{Debajyoti Choudhury \orcidlink{0000-0002-8124-0043},}
\emailAdd{debchou.physics@gmail.com}
\author[a]{Vineet K. Jha \orcidlink{0009-0000-2605-2544},}
\emailAdd{vineet.phd2022@physics.du.ac.in}
\author[b]{Rameswar Sahu \orcidlink{0000-0001-6587-951X},}
\emailAdd{rameswarsahu1@gmail.com}
\affiliation[a]{Department of Physics and Astrophysics, University of Delhi,
  Delhi 110 007, India.}
\affiliation[b]{Department of Physics, Indian Institute of Technology Kanpur, Uttar Pradesh 208016, India.}  

\abstract{A lowering of the post-inflation reheating temperature
  $T_{\rm RH}$ has attracted much attention as a means to facilitate
  obtaining the correct Dark Matter (DM) relic abundance. We take a
  relook at the paradigm under the assumption that the inflaton decay
  is essentially instantaneous with the thermal bath coming into
  existence only at a relatively low temperature. We demonstrate that,
  contrary to the popular misconception, the relevant scale is not
  necessarily the DM mass itself, but the characteristic temperature
  at which the dominant DM production processes are thermally
  efficient. With $T_{\rm RH}$ setting the maximum temperature
  available for thermal production, and thereby determining the
  accessible stages of the production history, lowering $T_{\rm
    RH}$ tends to reduce the relic abundance. While this effect could,
  in principle, be offset by altering the couplings defining the
  theory, there exists a minimum $T_{\rm RH}$---determined by the mass
  spectrum---below which the relic abundance falls to minuscule levels,
  whether the mechanism be freeze-in or freeze-out.}
\maketitle
\flushbottom
\section{Introduction}
\label{sec:intro}
Despite accounting for approximately $85\%$ of the total matter
content of the Universe \cite{Planck:2018vyg}, the fundamental nature
of Dark Matter (DM) remains a major open question. A wealth of
astrophysical and cosmological observations provide compelling
evidence for its existence \cite{Rubin:1970zza, Fixsen:1996nj,
  Planck:2018vyg, Primack:1997av, Clowe:2003tk}, yet its microscopic
properties, including its mass, spin, and possible interactions with
the Standard Model (SM) particles, remain largely unconstrained
(see, {\em e.g.}, refs.~\cite{ Arcadi:2017kky,
  Arcadi:2024ukq, Panci:2024oqc, Bernal:2017kxu, Bertone:2004pz} for
some recent reviews). Over the past several decades, extensive
experimental efforts have therefore been devoted to probing possible
non-gravitational interactions between DM and the SM through
direct-detection experiments~\cite{Misiaszek:2023sxe,
  Schumann:2019eaa, Akerib:2024ocy, XENON:2023cxc, XENON:2024wpa,
  PandaX:2024pjr}, indirect searches~\cite{Conrad:2017pms,
  Wood:2015ofa, Fermi-LAT:2015att, HESS:2018kom, IceCube:2024yaw,
  IceCube:2023ies}, and collider experiments~\cite{CMS:2024zqs,
  DeRoeck:2024fjq, PerezAdan:2023rsl}. Despite the remarkable progress
and increasing sensitivity of these searches, no conclusive evidence
for such interactions has been established to date. Similarly, the
very mechanism responsible for the establishment of the DM relic
density remains uncertain. Indeed, even whether DM was once
in thermal equilibrium with the bath before decoupling from it (the
\emph{freeze-out} mechanism \cite{Gondolo:1990dk}), or whether it was
produced gradually from the thermal bath during the evolution of the
Universe (the \emph{freeze-in} mechanism \cite{Hall:2009bx}), is still
entirely unknown.

The calculation of the DM relic abundance is
conventionally carried out under the assumption of a standard
cosmological history: the inflaton decays instantaneously, the thermal
bath of SM particles begins its evolution at a high temperature, and
the Universe remains radiation-dominated throughout the era between
reheating and Big Bang Nucleosynthesis (BBN). Whether it be the
freeze-out of thermally coupled DM or the freeze-in of feebly coupled
DM, either is ordinarily assumed to take place entirely within this
radiation-dominated epoch. However, the thermal history of the
Universe prior to BBN is only weakly constrained observationally, and
several well-motivated deviations from this picture have been studied
in the literature \cite{Giudice:2000ex, Fornengo:2002db,
  Pallis:2004yy, Gelmini:2006pw, Drees:2006vh, Yaguna:2011ei,
  Roszkowski:2014lga, Drees:2017iod, Bernal:2018ins, Bernal:2018kcw,
  Cosme:2020mck, Arias:2021rer, Bernal:2022wck, Bhattiprolu:2022sdd,
  Haque:2023yra, Ghosh:2023tyz, Silva-Malpartida:2023yks,
  Arias:2023wyg, Bernal:2024yhu, Silva-Malpartida:2024emu,
  Bernal:2024ndy, Roy:2025moo}. If, instead, the inflaton decay width
is very small, reheating is prolonged and may be accompanied or even
preceded by an epoch of matter domination; continuous entropy
injection from inflaton decay \cite{Allahverdi:2020bys,
  Batell:2024dsi} during this epoch can dilute a dark matter species
that freezes out \cite{Gelmini:2006pw, Bernal:2022wck, Bernal:2024yhu}
or freezes in \cite{Silva-Malpartida:2023yks,
  Silva-Malpartida:2024emu} before reheating completes, generically
suppressing its relic abundance relative to the standard prediction
for the same model parameters. Non-standard equations of state, such
as a kination-dominated epoch~\cite{Zeldovich:1961sbr, Spokoiny:1993kt,
  Joyce:1996cp, Ferreira:1997hj}, can similarly alter the relic
density prediction \cite{Belanger:2025ack}. 

In this work, we investigate
a different, and equally plausible, deviation: the inflaton decay
remains instantaneous and no entropy is ever injected into the SM bath
after reheating, but the bath itself is assumed to come into existence
only at a relatively low temperature $T_{\rm RH}$, taken here to lie
between a few GeV and several tens of TeV, rather than at the very
high temperatures that are usually assumed \cite{Bhattiprolu:2022sdd,
  Harigaya:2014waa,Cosme:2023xpa,Boddy:2024vgt}. We study how this
single change, {\em viz.} a low but otherwise standard (\emph{i.e.},
instantaneous production of the particles and creation of the thermal
bath) reheating temperature modifies the relic density estimate for
both the freeze-out and the freeze-in mechanisms.

The effect of a low $T_{\rm RH}$ on
dark-matter production has been explored in several contexts,
including scenarios in which the maximum temperature of the thermal
bath falls below the mass scale of the dark matter and production from
thermal annihilations becomes strongly suppressed~\cite{Boddy:2024vgt,
  Bhattiprolu:2022sdd, Cosme:2023xpa}. Here we emphasize a somewhat
broader point: the relevant scale is not necessarily the DM mass
itself, but the characteristic temperature at which the dominant
production processes are thermally efficient. Consequently, a substantial
suppression of the relic abundance can arise even while $T_{\rm RH}$
remains above $m_\chi$. What matters is not simply whether the DM is
kinematically accessible, but how much of the thermal phase space
supporting the dominant production channel survives the finite upper
boundary of the thermal history. We investigate this interplay in a
simplified dark sector containing a stable Dirac fermion $\chi$ and a
massive vector mediator $V_\mu$, coupled to both the SM fermions and
the dark sector. We assume negligible direct production of the dark
sector from the inflaton, so that its abundance is generated entirely
through interactions with the thermal bath. A sufficiently large DM
production through the inflaton decay can lead to the early
thermalization of DM, making a genuine freeze-in scenario difficult to
realize. While establishing the existence of such models is not the
primary purpose of this work---and we remain largely agnostic about
their underlying details---we will show that such a scenario can indeed
be realized in principle.

This setup provides a particularly transparent laboratory for
following the redistribution of the DM production history as $T_{\rm
  RH}$ is lowered. In the freeze-in regime, the abundance can be
generated through mediator decays as well as through annihilation
processes involving SM fermions and the mediator. Although the
relative importance of these channels is fixed, in the high-reheating
limit, for a given set of particle-physics parameters, altering the
thermal history can change their hierarchy. In particular, even when
the mediator decay is accessible and is the dominant production mode
when the reheating temperature is high, a low reheating temperature
instead can suppress the on-shell mediator production and can expose
an intermediate temperature regime in which off-shell annihilation
becomes the dominant source of DM. The freeze-in analysis therefore
goes beyond a simple suppression of the final yield: it reveals how a
finite maximum temperature can reshape the production history itself.

We also examine the corresponding freeze-out regime, not as a separate
source of novel thermal-history effects, but to complete the
connection between the two mechanisms of thermal relic
production. Once chemical equilibrium is established, the subsequent
freeze-out evolution largely erases the memory of the initial
temperature; deviations arise only when $T_{\rm RH}$
is low enough to interfere with thermalization. Following the
coupling required to reproduce the observed relic abundance across
both regimes then leads to a further consequence. As $T_{\rm RH}$ is
lowered, the feeble coupling required for freeze-in increases until it
approaches the boundary of thermalization, while the freeze-out
solution is reached from the opposite side. Below a model-dependent
minimum reheating temperature, neither side can reproduce the observed
abundance: weaker couplings yield insufficient freeze-in production,
whereas stronger couplings lead to excessive annihilation after
thermalization. Within the assumed reheating history, this establishes
a direct link between the particle-physics parameters of a dark-sector
realization and the minimum temperature at which the post-inflationary
thermal bath must have been established, providing a possible handle
on the otherwise poorly constrained pre-BBN cosmological history.

The remainder of this paper is organized as follows. In
Section~\ref{sec:relic}, we review the calculation of the dark matter
relic abundance under both standard and low-reheating cosmologies. In
Section~\ref{sec:model}, we introduce the simplified dark matter model
considered in this work. The resulting phenomenology and relic-density
predictions are presented in Section~\ref{sec:pheno}. Finally, we
summarize our results and conclude in Section~\ref{sec:sum}.



\section{Relic-density estimation}
\label{sec:relic}

Turning to the calculation of the relic density itself, we first
recall the standard Boltzmann treatment of freeze-out and freeze-in,
common to any radiation-dominated history, and then identify precisely
which elements of this treatment are altered once the SM bath is assumed to originate at a
relatively low $T_{\rm RH}$. Since reheating is assumed to be
instantaneous, the Universe is radiation dominated for all $T<T_{\rm
  RH}$, and no subsequent entropy injection from the inflaton sector
is present. The expansion rate and the entropy
density, therefore, retain their standard
radiation-dominated forms, namely
\begin{equation} \label{eq:stdHs}
H(T)=\sqrt{\frac{\pi^2 g_*(T)}{90}}
\frac{T^2}{M_{\rm Pl}},
\qquad
s(T)=\frac{2\pi^2}{45}g_{*s}(T)T^3,
\end{equation}
where $g_*$ and $g_{*s}$ denote the effective numbers of relativistic
degrees of freedom associated with the energy and entropy densities
respectively, and $M_{\rm Pl}$ is the reduced Planck mass. Entropy
conservation after reheating implies $s a^3={\rm const}$, where
  $a$ denotes the instantaneous scale factor of the assumed
  Friedmann-Lema\^itre-Robertson-Walker universe. Consequently, the
effect considered here does not arise from a modified expansion rate
or a late-time dilution of the DM abundance. Instead, $T_{\rm RH}$
sets the highest temperature available to the thermal bath and,
therefore, provides the initial condition for DM evolution.

For a DM species $\chi$ whose number density is controlled by
annihilation and inverse-annihilation processes involving particles in
the thermal bath, its evolution can be expressed in terms of the
comoving abundance ($Y_\chi=n_\chi/$s). Introducing
\begin{equation}
x\equiv\frac{m_\chi}{T},
\qquad
x_{\rm RH}\equiv\frac{m_\chi}{T_{\rm RH}},
\end{equation}
the Boltzmann equation takes the standard form
\begin{equation}
\frac{dY_\chi}{dx}
 \approx
\frac{s(T)\langle\sigma v\rangle}{xH(T)}
\left[
Y_{\chi,{\rm eq}}^2-Y_\chi^2
\right],
\label{eq:boltzmann_general}
\end{equation}
up to the usual correction associated with the temperature dependence
of $g_{*s}$. The distinction from the conventional high-temperature
treatment is that Eq.~\eqref{eq:boltzmann_general} is to be
evolved only from $x=x_{\rm RH}$, rather than from an arbitrarily
small value of $x$.

The relevance of this finite lower boundary depends qualitatively on
whether the DM abundance is generated through freeze-in or freeze-out.

\subsection{Freeze-in}

For freeze-in production, the interaction between the dark and visible
sectors is sufficiently weak so that $\chi$ never reaches thermal
equilibrium. Accordingly,
\begin{equation}
Y_\chi\ll Y_{\chi,{\rm eq}},
\end{equation}
and the back reaction proportional to $Y_\chi^2$ can be
neglected\footnote{This and other approximations in this section
  are only in the spirit of aiding analytic solutions so as to afford
  an easy understanding. For all our numerical work, the exact
  equations are used in each case.}. Equation~\eqref{eq:boltzmann_general} then
reduces to
\begin{equation}
\frac{dY_\chi}{dx}
\simeq
\frac{s(T)\langle\sigma v\rangle}{xH(T)}
Y_{\chi,{\rm eq}}^2.
\label{eq:freezein_boltzmann}
\end{equation}
Assuming a negligible initial DM abundance at reheating, namely,
\begin{equation}
Y_\chi(x_{\rm RH})=0,
\end{equation}
the asymptotic abundance is given by
\begin{equation}
Y_\chi^\infty(T_{\rm RH})
\simeq
\int_{x_{\rm RH}}^\infty
dx
\frac{s(T)\langle\sigma v\rangle}{xH(T)}
Y_{\chi,{\rm eq}}^2.
\label{eq:freezein_integral}
\end{equation}
The dependence on $T_{\rm RH}$ is therefore explicit: lowering the
reheating temperature increases the lower integration boundary $x_{\rm
  RH}$ and removes the contribution from temperature regimes $T>T_{\rm RH}$. Equivalently, writing the
production in terms of a reaction density $\gamma_\chi(T)\equiv n_{\rm
  eq}^2 \langle \sigma v\rangle$,
\begin{equation}
Y_\chi^\infty(T_{\rm RH})
 \simeq
\int_{0}^{T_{\rm RH}}
\frac{dT}{sHT}
\gamma_\chi(T),
\label{eq:freezein_temperature}
\end{equation}
where the mild corrections arising from the temperature dependence of
the effective degrees of freedom are again understood.

Equation~\eqref{eq:freezein_temperature} makes the physical role of
$T_{\rm RH}$ transparent. The sensitivity of the relic abundance to
reheating is determined by the temperature dependence of the
production rate. If the dominant contribution to the integral arises
well below $T_{\rm RH}$, reducing $T_{\rm RH}$ has little
effect. However, once $T_{\rm RH}$ approaches or falls below
the temperature range in which the production kernel
$\gamma_\chi/(sHT)$ is largest, the corresponding
contribution is progressively removed and the final abundance
decreases. The resulting dependence need not follow a universal power
of $T_{\rm RH}$; rather, it is fixed by the underlying
production channels, masses, thresholds, and resonant structures of
the model.

\subsection{Freeze-out}

The role of $T_{\rm RH}$ is qualitatively different for thermal
freeze-out. In the conventional scenario, DM is initially maintained
in chemical equilibrium,
\begin{equation}
Y_\chi\simeq Y_{\chi,{\rm eq}},
\end{equation}
and departs from equilibrium around a freeze-out temperature ($T_f$),
determined parametrically by
\begin{equation}
n_{\chi,{\rm eq}}(T_f)
\langle\sigma v\rangle
\sim H(T_f).
\label{eq:freezeout_condition}
\end{equation}
If $T_{\rm RH}\gg T_f$, and the interaction rate is sufficiently large
enough to establish chemical equilibrium after reheating, the
system loses memory of the initial temperature (even if it were only
moderately high) before freeze-out. The subsequent evolution is then
identical to the standard radiation-dominated calculation, and the
final relic abundance is essentially independent of $T_{\rm RH}$.

The situation changes when $T_{\rm RH}$ approaches the freeze-out
temperature. The Boltzmann evolution then essentially begins
only at
\begin{equation}
x_{\rm RH}=\frac{m_\chi}{T_{\rm RH}},
\end{equation}
and the high-temperature part of the thermal history is absent. Whether a conventional freeze-out description remains applicable must then be determined dynamically by comparing the interaction rate with the Hubble expansion rate after reheating. In particular, the condition
\begin{equation}
\Gamma_\chi(T)
\equiv
n_{\chi,{\rm eq}}(T)\langle\sigma v\rangle
\gtrsim H(T)
\end{equation}
must be satisfied over a sufficient temperature interval below $T_{\rm
  RH}$ for $\chi$ to attain chemical equilibrium. If this occurs, the
subsequent decoupling can still be treated as thermal freeze-out,
albeit with a potentially shortened equilibrium epoch. If it does not,
imposing $Y_\chi(T_{\rm RH})=Y_{\chi,{\rm eq}}(T_{\rm RH})$ is not
justified, and the abundance becomes sensitive to the initial
condition at reheating. In this regime, the production history
continuously approaches a non-equilibrium freeze-in-like evolution
rather than an ordinary freeze-out solution.


When $T_{\rm RH}\ll T_f$, the DM cannot attain chemical equilibrium
after reheating. In this limit, $Y\ll Y_{\rm eq}$ throughout the
relevant evolution and, consequently,
Eq.~\eqref{eq:boltzmann_general} reduces to the
freeze-in-like equation, Eq.~\eqref{eq:freezein_boltzmann}. The
asymptotic yield is therefore determined by the production occurring
after reheating and is given by Eq.~\eqref{eq:freezein_integral}. For
a standard radiation-dominated Universe, with $H(T)$ and $s(T)$ given
by Eq.~\eqref{eq:stdHs}, and assuming
temperature-independent\footnote{This approximation is justified when
$g_*$ and $g_{*S}$ vary negligibly over the relevant temperature
range, which is the case when no mass threshold is crossed. The
thermally averaged annihilation cross section is, likewise,
approximately temperature independent when the annihilation is
dominated by the $s$-wave contribution.} $g_*$, $g_{*S}$, and
$\langle\sigma v\rangle$, the resulting yield can be written as
\begin{equation}
Y_\infty(x_{\rm RH})
= C \int_{x_{\rm RH}}^\infty xe^{-2x}dx,
\label{eq:yield3}
\end{equation}
where
\begin{equation}
C =
\frac{M_{\rm Pl}}{1.66\sqrt{g_*}}
\left(\frac{45g_\chi^2}{2\pi^2g_{*s}}\right)
\frac{m_\chi}{(2\pi)^3}
\langle\sigma v\rangle \ ,
\end{equation}
and we have used the nonrelativistic form of the equilibrium yield, namely,
\begin{equation}
Y_{\rm eq}\equiv\frac{n_{\rm eq}}{s}
\sim
\left(\frac{45g_\chi}{2\pi^2g_{*S}}\right)
\left(\frac{x}{2\pi}\right)^{3/2}e^{-x} \ .
\end{equation}
The integral then gives
\begin{equation}
Y_\infty(x_{\rm RH})
 \simeq C\left(1+2x_{\rm RH}\right)e^{-2x_{\rm RH}}.
\label{eq:yield4}
\end{equation}
This result makes the dependence on the reheating temperature explicit. Since $x_{\rm RH}=m_\chi/T_{\rm RH}$, the relic abundance exhibits an exponential suppression, $\Omega_\chi h^2\propto(1+2m_\chi/T_{\rm RH})$ $e^{-2m_\chi/T_{\rm RH}}$, as $T_{\rm RH}$ is lowered below the freeze-out temperature. 

Thus, within the cosmology considered here, the effect of $T_{\rm RH}$
can be understood as a finite-temperature boundary condition on an
otherwise standard radiation-dominated Boltzmann evolution. For
freeze-in, the finite value of $T_{\rm RH}$ directly truncates the
production integral and can therefore strongly suppress the final
abundance. For freeze-out, in contrast, the standard
prediction is recovered whenever the reheating temperature is
sufficiently high for chemical equilibrium to be established well
before decoupling; appreciable deviations arise only when the finite
thermal history interferes with the equilibration and freeze-out
dynamics. The present-day relic density is finally obtained from the
asymptotic yield through
\begin{equation}
\Omega h^2 = \frac{m_\chi s_0 Y_\chi^\infty} {\rho_c/h^2},
\label{eq:relic_density}
\end{equation}
where $s_0$ and $\rho_c$ denote the present entropy density and
the critical density, respectively.

\section{Model}
\label{sec:model}

The purpose of this paper is to illustrate the generic behavior of
dark matter production in scenarios with low reheating
temperatures. To this end, we consider a highly simplified setup
consisting of a stable dark Dirac fermion $\chi$ (which serves as our
DM candidate), a heavy vector mediator $V$, and the SM degrees of
freedom.  We assume that the DM, as well as some of the SM fermions,
are charged under the gauge group $U(1)_V$, thereby facilitating the
DM's interactions with the visible sector.  $V$ subsequently acquires
a mass through a \emph{St\"uckelberg-like} mechanism\footnote{While
the usual route of a Higgs field could be adopted as well, we desist
from introducing an additional particle in our discussions.}. In this
setup, the dark-sector Lagrangian is given by
\begin{equation}
\label{eq:lag}
\mathcal{L} \supset
\bar{\chi}\left(i\slashed{D}-m_\chi\right)\chi
-\frac{1}{4}V_{\mu\nu}V^{\mu\nu}
+\frac{1}{2}m_V^2V^\mu V_\mu \ ,
\end{equation}
with the field-strength tensor and the covariant derivative defined as
\begin{equation*}
V_{\mu\nu}=\partial_\mu V_\nu-\partial_\nu V_\mu, \qquad
D_\mu\chi=\left(\partial_\mu-i g_\chi V_\mu\right)\chi .
\end{equation*}
While $V$ could, in principle, couple with all SM fermions,
constraints exist in the $(g_f, m_V)$ planes---$g_f$ being its
coupling with a typical fermion---from both low-energy experiments as
well as non-observation at colliders. The simplest way to evade most
constraints is to postulate that $V$ has purely vector couplings to
the SM fermions and that only the third generation fermions carry a
charge under it. This leads to
\begin{equation}
\mathcal{L}\supset
-\sum_i g_{f_i} V_\mu\,\bar{f_i}\gamma^\mu f_i ,
\end{equation}
where the sum extends over $t, b, \tau$ and $\nu_\tau$. An
anomaly-free construction demands that the only non-zero couplings
satisfy
\begin{equation}
  g_f \equiv g_t = g_b = \frac{-g_\tau}{3} = \frac{-g_{\nu_\tau}}{3} \ ,
\label{eq:sm_coupling}
\end{equation}
To reduce the number of parameters, for the rest of the analysis, we
hold\footnote{Note that $g_f$ and $g_\chi$ are absolutely independent
of each other and could even differ by several orders of magnitude, as
can occur in extra-dimensional theories.} $g_f = 0.1$ while leaving
$g_\chi$ free.

While more complicated solutions are indeed possible (such as a $g_f$
that is generation dependent and suitably suppressed, but non-zero,
for the first two generations), with interesting ramifications, the
simplistic construction here serves to illustrate our point, while
being phenomenologically consistent\footnote{It is important to note
that the SM Yukawa sector remains unmolested except for terms that
connect the third generation to the first two. This could be
addressed, for example, by invoking a Froggatt-Nielsen-like
mechanism. We, however, desist from doing so as the flavour problem is
not of interest to us.}.

In this work, we assume that the inflaton decays predominantly into SM
particles, while its decay into the dark sector is
negligible. Consequently, the dark sector is initially absent, or at
most highly suppressed, and is subsequently populated through its
interactions with the SM bath. This assumption is important for
maintaining a genuine freeze-in regime: sizable direct inflaton
couplings to the dark sector could, instead, produce a non-negligible
DM abundance during reheating and, depending on the interaction
strength (which can be as large as $\mathcal{O}(1)$ in our model),
potentially could bring the dark sector into equilibrium at an early
stage.

Such a hierarchy between the inflaton ($\phi$) couplings to the SM and
the dark sector can be realized in a
  multitude of models. An example is afforded by extra-dimensional
scenarios \cite{Cox:2012ee,Giudice:2016yja,Giudice:2017fmj} with
spatially separated branes, wherein one may
localize the SM fields and the inflaton on one brane, while the DM is
localized on a different brane and the vector mediator $V$ propagates
in the bulk. The bulk vector can contain a massless mode together with
a tower of heavier (Kaluza-Klein) modes. The latter can be integrated
out in the decoupling limit,
$m_{n>1} \rightarrow \infty$, leaving the
massless mode as the relevant low-energy mediator. The massless mode,
corresponding to the unbroken $U(1)$ group, can subsequently acquire a
mass, $m_V$, through the \emph{Stückelberg-like} mechanism\footnote{A
similar kind of mass generation has been studied in
ref.~\cite{Lee:2017fin}.}, as we mentioned earlier.

The inflaton can decay directly into SM particles through various
types of interactions, the simplest one being $\mu \phi H^\dagger H$,
where $H$ is the SM Higgs and $\mu$ a mass-dimension-one coupling. When kinematically allowed, this would engender
  several decays of the form $\phi \to H_i H_i$ (with no summation over $i$), where $H_i$ denotes the
corresponding Higgs component or scalar mode. The identification
  of the modes depends on whether the decay occurs below or above the
  electroweak scale\footnote{Above the electroweak scale, $H_i$ denotes
the components of the $SU(2)_L$ Higgs doublet, while below the
electroweak scale it denotes the corresponding scalar mass eigenstates
after electroweak symmetry breaking. At finite temperature, the masses
of these states should strictly be understood as temperature-dependent
effective (thermal) masses.}.  Taking into consideration that the masses
  receive both usual renormalization as well as thermal corrections,
  the corresponding decay width can be notionally approximated
  by\footnote{Actually, we have $\mu \equiv \mu(T)$, and the
corresponding thermal corrections should, in principle, be taken into
account. However, we neglect these corrections for the purpose of
obtaining a straightforward estimate without losing the qualitative
aspects.}
\[  \Gamma_{\phi \rightarrow H_i H_i} \propto \frac{\mu^2}{32 \pi m_\phi} \sqrt{1-\frac{4 m_H^2}{m_\phi^2}} ,\]
with $m_{\phi/H}$ denoting the masses of the
inflaton and the Higgs component respectively. This can easily be
large enough for essentially instantaneous reheating. For $m_\phi < 2
m_H$, the prominent inflaton tree-level decay modes would be $\phi \to
H + \psi \bar \psi$ through an off-shell $H$ (equivalently, through a
dimension-five operator of the form $(y / \Lambda) \phi H \bar \Psi_L
\psi_R$, where $\psi$ denotes the SM fermion species of all
generations, and $\Psi_L, \psi_R$ are the $SU(2)$ doublet (singlet)
fermion fields, $y$ is a dimensionless Yukawa-like coupling and
$\Lambda$ is the scale of suppression).

On the other hand, the inflaton is postulated to have no direct local
interaction with the DM on account of them being localized on
different branes. The interaction of the DM with the SM sector and the
inflaton can proceed only through the vector mediator\footnote{Although
gravitational interactions and, in some cases, interactions mediated
by model-dependent extra fields such as the dilaton, radion, etc., may
exist, these interactions are, typically, further suppressed and can
therefore be safely neglected.}. Since the inflaton is, presumably, a
real scalar, the dominant interaction with $V^\mu$ would be of the
form $\phi V_{\mu\nu}V^{\mu\nu} / \Lambda$ generated through the heavy SM fermion triangle loop. Consequently, the
  resulting production of the vector, if kinematically allowed, remains negligible, with $\Gamma_{\phi \rightarrow VV} \propto g_f^4
  y^2/\Lambda^2$. This, in turn, also ensures that
  DM production from inflaton decay 
  remains negligible, even for $g_\chi \sim \mathcal{O}(1)$.

We have implemented the model in
\texttt{FeynRules}~\cite{Alloul:2013bka} and generated the
corresponding \texttt{CalcHEP}~\cite{Belyaev:2012qa} model files. The
relic abundance is then calculated using
\texttt{micrOMEGAs}~6.2.3~\cite{Alguero:2023zol, Belanger:2026asz},
employing its \texttt{darkOmegaInfl} routine. This routine solves the
Boltzmann evolution starting from a vanishing dark-sector abundance at
the end of inflation and accounts for the reheating dynamics through
the inflaton decay width and the inflationary scale $H_I$
\cite{Belanger:2024yoj}. To realize the instantaneous-reheating limit
considered in this work, we choose the inflaton decay width
to be sufficiently large such that the reheating temperature $T_{\rm RH}$
and the maximum temperature $T_{\max}$ of the thermal bath coincide to
within numerical precision. We have explicitly verified this condition
for each of the benchmark points considered in our analysis. The reheating
temperature is then varied by varying $H_I$, while maintaining the
inflaton decay width to be in the instantaneous-reheating
regime. In this
limit, the thermal bath is effectively established at $T_{\rm RH}
\simeq T_{\max}$, which therefore sets the initial temperature for the
subsequent evolution of the SM bath and the dark sector. Without
  loss of generality, we take $T_{\rm RH}$ as the independent
  parameter throughout the remainder of this work. The relic-density
results obtained with \texttt{micrOMEGAs} are additionally
cross-checked against an independent numerical implementation of the
corresponding Boltzmann equations.

The primary objective of this work is to elucidate the characteristic
features of dark matter production in the low-reheating cosmology
discussed earlier, rather than to provide a comprehensive
phenomenological study of the particular simplified model introduced
above. The model is chosen as a minimal and transparent framework in
which the impact of a finite reheating temperature on both freeze-in
and freeze-out production can be isolated and studied in a controlled
manner. While the quantitative results necessarily depend on the
particle-physics realization, the underlying effects associated with
the finite upper boundary of the thermal history are not specific to
any particular model and are expected to arise more generally in
scenarios with analogous production mechanisms. A detailed analysis of
the experimental constraints on the model, including those from
direct-detection and collider searches is, therefore, beyond the scope
of this work. Rather, we take care that
  the parameter space (especially the benchmark scenarios) considered
  in our study is consistent with current phenomenological
  constraints. This choice allows us to disentangle the effects of
the low-reheating cosmology from model-dependent experimental
constraints and to provide a more transparent assessment of its impact
on the relic abundance. For a detailed discussion of the
direct-detection and collider constraints on this simplified dark
matter model, we refer the reader to
Refs.~\cite{Belanger:2024yoj,CMS:2021ctt,CMS:2021far,ATLAS:2017fih,ATLAS:2021kxv}.

\section{Phenomenology}
\label{sec:pheno}

Having outlined how a relatively low reheating temperature
is expected to modify either freeze-in or freeze-out
evolutions, we now investigate its phenomenological implications
considering, as a template,
the model introduced in Sec.~\ref{sec:model}. Although our
numerical analysis is performed for this specific model, the
underlying physical picture is expected to be generic to a broad class
of BSM scenarios in which the DM relic abundance is determined through
thermal freeze-in or freeze-out.

We first study the dependence of the freeze-in relic abundance on
$T_{\rm RH}$ for scenarios with kinematically forbidden and allowed
mediator decays. We then examine the corresponding freeze-out
abundance (albeit with different values of the coupling constant
$g_\chi$) and identify the range of reheating temperatures for which
the standard freeze-out picture remains applicable. Finally, we
determine the DM coupling required to reproduce the observed relic
abundance as a function of $T_{\rm RH}$, thereby illustrating the
transition between the freeze-in and freeze-out regimes. To reduce the
number of free parameters, the magnitude of the coupling
$g_f$ ---see Eq.~(\ref{eq:sm_coupling})--- is held fixed at 0.1
throughout.

\subsection{Freeze-In}

Within the present model, the DM abundance is primarily generated
through the following processes:
\begin{align}
V &\rightarrow \bar{\chi}\chi, \label{eq:decvxx}\\
VV &\rightarrow \bar{\chi}\chi, \label{eq:annvvxx}\\
\bar{f}f &\rightarrow \bar{\chi}\chi. \label{eq:annffxx}
\end{align}

The relative importances of these production channels depend on the
masses and couplings of the dark and visible sectors. When the decay
$V\rightarrow\bar{\chi}\chi$ is kinematically allowed, it typically
provides the dominant contribution to the freeze-in yield. This
follows from the parametric dependence of the corresponding rates: the
decay width scales as $\Gamma(V\rightarrow\bar{\chi}\chi)\propto
g_\chi^2$, whereas the annihilation processes are suppressed not only
by additional powers of the couplings, with
$\sigma(VV\rightarrow\bar{\chi}\chi)\propto g_\chi^4$ and
$\sigma(\bar{f}f\rightarrow\bar{\chi}\chi)\propto g_f^2g_\chi^2$, but
also on account of the flux factor. In other words, whenever the decay
channel is available, and the abundance of the vector is sufficient,
it generally dominates the DM production and determines the resulting
freeze-in abundance.

\begin{figure}[t]
\centering
\includegraphics[width = 2.8in]{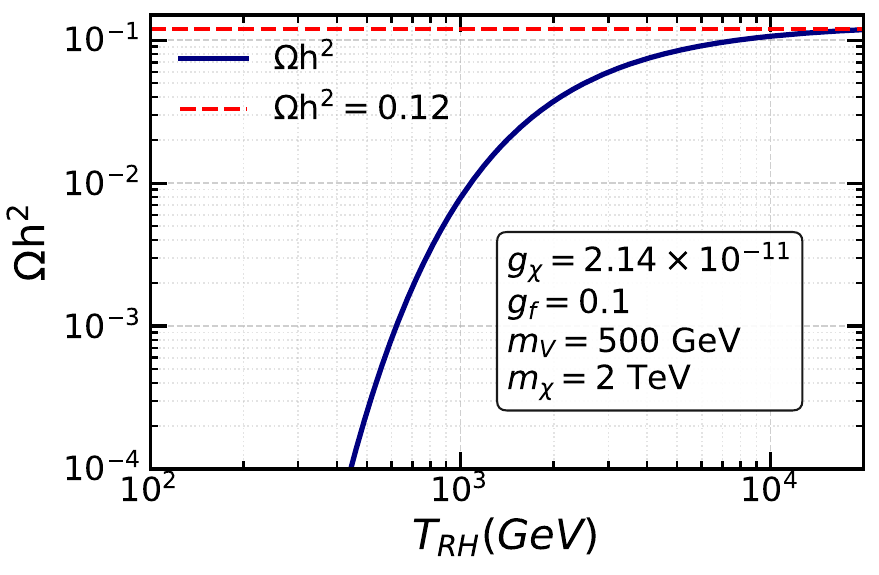} \quad
\includegraphics[width = 2.8in]{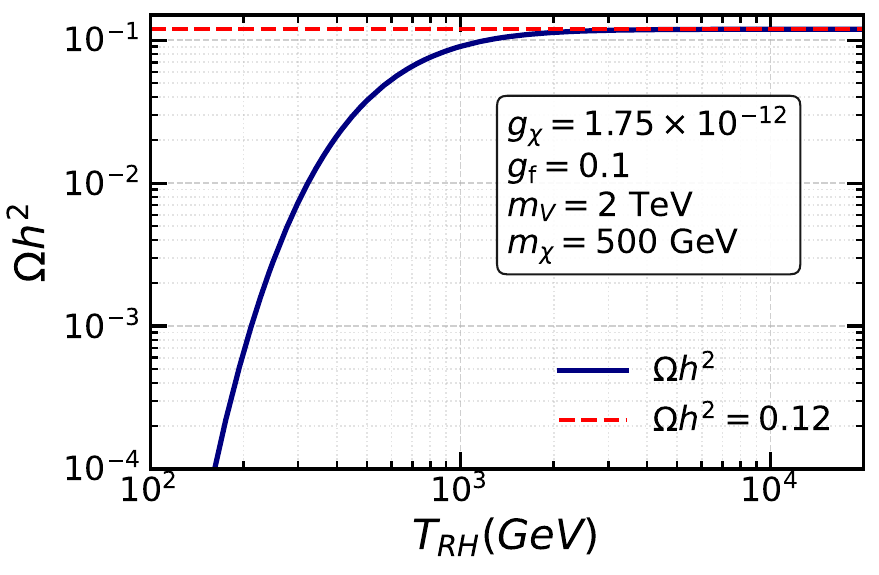}
\caption{Variation of the relic abundance, $\Omega h^2$, with the
  reheating temperature, $T_{\rm RH}$, for two benchmark scenarios:
  (a) $m_V=500~{\rm GeV}$, $m_\chi=2~{\rm TeV}$, $g_f=0.1$, and (b)
  $m_V=2~{\rm TeV}$, $m_\chi=500~{\rm GeV}$, $g_f=0.1$, corresponding
  to kinematically forbidden ($m_V<2m_\chi$) and allowed
  ($m_V>2m_\chi$) decay channels, respectively. In both cases, when
  $T_{\rm RH}$ is sufficiently larger than the relevant particle
  masses, the relic abundance approaches an asymptotic value that
  reproduces the observed dark matter abundance for
  $g_\chi=2.14\times10^{-11}$ and $g_\chi=1.75\times10^{-12}$,
  respectively.}
\label{plot:frz-in}
\end{figure}

We begin with the opposite case, namely, that of $V$
decay being kinematically forbidden. As a benchmark point, we choose
$m_V=500~\mathrm{GeV}$ and $m_\chi=2~\mathrm{TeV}$, for which the
observed relic abundance in the high-reheating temperature limit is
obtained only for an expectedly suppressed DM coupling
($g_\chi=2.14\times10^{-11}$) (see the left panel of
Fig.~\ref{plot:frz-in}). DM production proceeds entirely through the
annihilation channels, with $\bar{f}f\rightarrow
V^*\rightarrow\chi\bar{\chi}$ providing the dominant contribution for
this benchmark point. At a sufficiently high $T_{\rm RH}$, the thermal
bath contains an adequate population of energetic SM fermions to
sustain this production, and the relic abundance consequently
approaches its standard high-reheating value. Holding the parameters
$m_V$, $m_\chi$, and $g_\chi$ to the aforementioned values, as $T_{\rm RH}$ is
lowered, the available thermal energy becomes progressively
insufficient to efficiently produce the $\chi\bar{\chi}$ pair. Once
$T_{\rm RH}$ falls below the characteristic scale $2m_\chi$, the
relevant high-energy tail of the thermal distribution is exponentially
depleted, resulting in a precipitous suppression of the production
rate and, consequently, of the relic abundance. For this particular
benchmark point, this suppression drives the relic abundance down to $\Omega h^2\sim 10^{-4} \, (10^{-20})$ for $T_{\rm RH}\sim500 \, (100)~\mathrm{GeV}$.

The situation becomes considerably richer when the decay
$V\rightarrow\bar{\chi}\chi$ is kinematically accessible.  It can now
compete with the annihilation channels, with the extent of the
competition determined by the maximum temperature of the thermal
bath. To illustrate this interplay, we consider four representative
benchmarks. The first (illustrated in the right panel of
Fig.~\ref{plot:frz-in}) is characterized by interchanging the masses,
namely now $m_V=2~\mathrm{TeV}$ and $m_\chi=500~\mathrm{GeV}$. To
reproduce the observed relic abundance in the
high-reheating-temperature limit, the DM coupling has to be lowered
further down to $g_\chi=1.75\times10^{-12}$. This benchmark point is
particularly interesting as, in addition to exhibiting the effect of a
finite reheating temperature, such a pairing of the masses also admits
a corresponding thermal freeze-out solution (albeit for a larger
$g_\chi$) allowing for a direct connection with the freeze-out
analysis presented in the following subsection. However, the
relatively modest hierarchy between $m_V$ and $m_\chi$ leaves only a
narrow temperature interval for the transition between the different
production mechanisms to be clearly resolved. Nonetheless, it is
interesting to note that the relic density exhibits a considerably
slower fall-off than in the left panel. This behaviour can be
understood by considering the different regimes of $T_{\rm RH}$. For
$T_{\rm RH}\sim {\cal O}({\rm TeV})$, DM production is dominated by
the decay channel, and lowering $T_{\rm RH}$ progressively suppresses
the equilibrium density $n_V$ of $V$. Although the decay contribution
is consequently reduced, it still remains significant over this
range and provides an additional source of DM production that was
absent for the benchmark point shown in the left panel. As
$T_{\rm RH}$ is lowered further, $n_V$ becomes sufficiently suppressed
for $\bar f f\to\bar\chi\chi$ to take over as the dominant production
mode, as in the left panel. Nevertheless, the relic density continues
to decrease more slowly than in the left panel. This can be attributed
to the comparatively smaller value of $m_\chi$, which allows the
high-energy tail of the thermal distribution to sustain DM production
down to lower values of $T_{\rm RH}$. For this particular benchmark,
however, the regime in which annihilation dominates DM production
overlaps substantially with the onset of thermal suppression of the
annihilation process itself, occurring for $T_{\rm RH}\lesssim
2m_\chi$. Consequently, the intermediate regime in which annihilation
dominates over decay while DM production remains thermally
unsuppressed is only marginally discernible.

\begin{figure}[t]
\centering
\includegraphics[width = 4.0in]{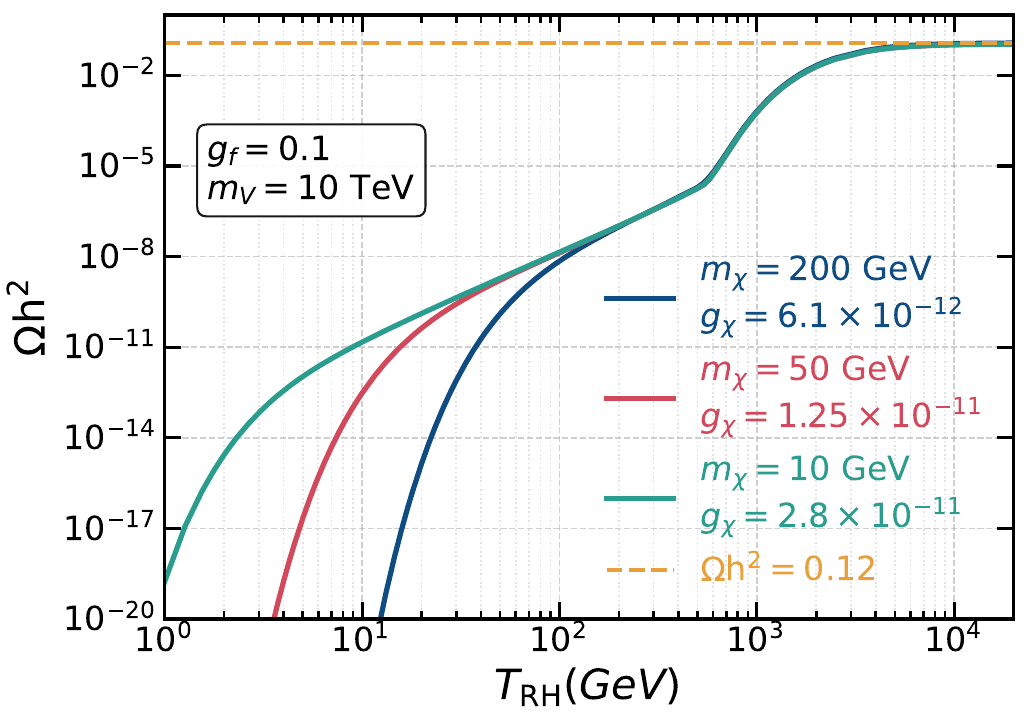}
\caption{Variation of the relic abundance, $\Omega h^2$, with the
  reheating temperature, $T_{\rm RH}$, illustrating the impact of the
  dominant annihilation process $f\bar{f}\rightarrow
  V^*\rightarrow\chi\bar{\chi}$ between two successive exponential
  suppressions. The mediator mass is fixed at $m_V=10~{\rm TeV}$, with
  its coupling to SM fermions fixed at $g_f=0.1$, while the DM mass is
  varied as $m_\chi=200~{\rm GeV}$ (blue), $50~{\rm GeV}$ (red), and
  $10~{\rm GeV}$ (green). A larger mass hierarchy between the mediator
  and DM allows a wider range of $T_{\rm RH}$ over which the
  contribution of this annihilation channel becomes more pronounced.}
\label{plot:frz-in_2}
\end{figure}

To illustrate this structure more distinctly, we consider three
additional benchmark points. While fixing $m_V=10~\mathrm{TeV}$, we
choose three representative values of the DM mass, {\em viz.}
$m_\chi=10$, $50$, and $200~\mathrm{GeV}$. Once again, in each case,
the value of $g_\chi$ is chosen such that the observed relic abundance
is reproduced in the high $T_{\rm RH}$ limit. The resulting
dependence of $\Omega h^2$ on $T_{\rm RH}$ is displayed in
Fig.~\ref{plot:frz-in_2}. The larger separation between the mediator
and DM mass scales provides a broader temperature interval in which
the different production mechanisms can be disentangled.
  
Several distinct regimes can be identified in
Fig.~\ref{plot:frz-in_2}. At a sufficiently high $T_{\rm RH}$, all
relevant production channels are thermally accessible (with the decay
channel dominating) and the relic abundance approaches its standard
high-reheating value. As $T_{\rm RH}$ is lowered below $m_V$, the
thermal population of $n_V$ of $V$ is depleted and the
corresponding DM yield undergoes an exponential suppression, marking
the first departure from the high-reheating regime. For
  $T_{\rm RH}\lesssim \mathcal{O}({\rm TeV})$,  DM
  production dominantly proceeds through 
  $\bar f f\rightarrow V^*\rightarrow\chi\bar\chi$. In the regime
  $\sqrt{s}\ll m_V$, the corresponding cross section scales as
\[\sigma(\bar f f\rightarrow\chi\bar\chi) \propto \frac{g_\chi^2 g_f^2}{m_V^4}s.\]
With $s\sim T^2$, the relic abundance scales as $\Omega_\chi
h^2\propto T_{\rm RH}^3$ for fixed $m_\chi$, $g_\chi$, $g_f$, and
$m_V$, which explains the cubic suppression with decreasing $T_{\rm
  RH}$ as shown in Fig.~\ref{plot:frz-in_2}. Unlike in the
previous case, this region is well above the masses of the DM and can
be clearly distinguished from the region where $T_{\rm RH}$ falls
below the characteristic scale $2m_\chi$, and the production of the DM
pair itself becomes strongly Boltzmann suppressed. The annihilation
contribution thereafter decreases exponentially, giving rise to a
second pronounced suppression of the relic abundance. The resulting
behavior therefore consists of two successive suppressions, associated
respectively with the depletion of on-shell mediator production and
the subsequent suppression of DM-pair production, separated by an
intermediate regime in which off-shell annihilation dominates.

This structure is most clearly resolved for the
$m_\chi=10~\mathrm{GeV}$ benchmark, for which the large hierarchy
$m_V/m_\chi$ provides a sufficiently broad separation between the two
characteristic temperature scales. As $m_\chi$ is increased while
keeping $m_V$ fixed, these scales move closer together and the
intermediate annihilation-dominated regime correspondingly
narrows. This is precisely why the separation was only barely
  discernible  for the $(m_\chi=500~\mathrm{GeV},
  m_V=2~\mathrm{TeV})$ benchmark point discussed
  earlier.


\subsection{Freeze-Out}
\label{ssec:fzout}

We now examine the impact of a relatively low $T_{RH}$ within the
thermal freeze-out paradigm. This provides a useful complement to the
freeze-in analysis and allows us to distinguish the effects arising
from the limited production temperature from those associated with the
establishment of chemical equilibrium. We consider the first two
benchmark scenarios introduced in the previous subsection, (a) $m_\chi
= 2~{\rm TeV}, m_V = 500~{\rm GeV}$ and (b) $m_\chi = 500~{\rm GeV},
m_V = 2~{\rm TeV}$, but now allow for much larger values for $g_\chi$.

The two benchmarks provide complementary realizations of the dominant
DM annihilation processes. For benchmark (a), the channel
$\chi\bar{\chi}\rightarrow VV$ is kinematically open and, together
with $\chi\bar{\chi}\rightarrow\bar f f$, contributes to the depletion
of the DM abundance. For benchmark (b), on the other hand, the
annihilation into a pair of on-shell $V$'s is highly suppressed, and
the depletion dominantly proceeds through $\chi\bar{\chi}\rightarrow
V^*\rightarrow\bar f f$. The latter, despite suffering a $s$-channel
suppression, remains sufficiently efficient (for the chosen
parameters) to reproduce the observed relic
abundance through thermal freeze-out. This is not the case for the
three additional benchmarks considered in the previous section, for
which the larger mediator mass leads to a stronger suppression of the
annihilation rate and, consequently, these benchmarks do not admit a
thermal freeze-out solution that reproduces the observed relic
abundance (unless, of course, the gauge couplings are increased to
nonperturbative levels).

\begin{figure}[!ht]
\subfloat[]{\includegraphics[width = 2.8in]{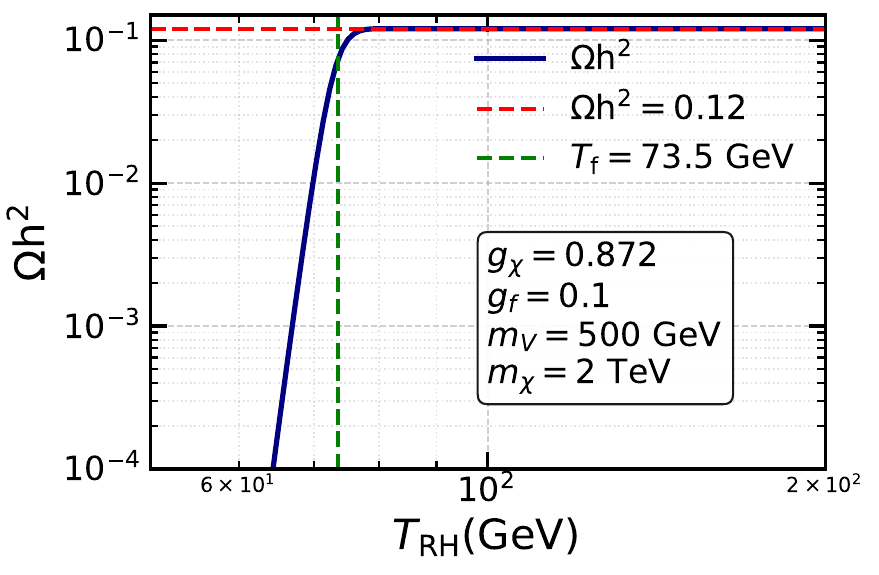}} \quad
\subfloat[]{\includegraphics[width = 2.8in]{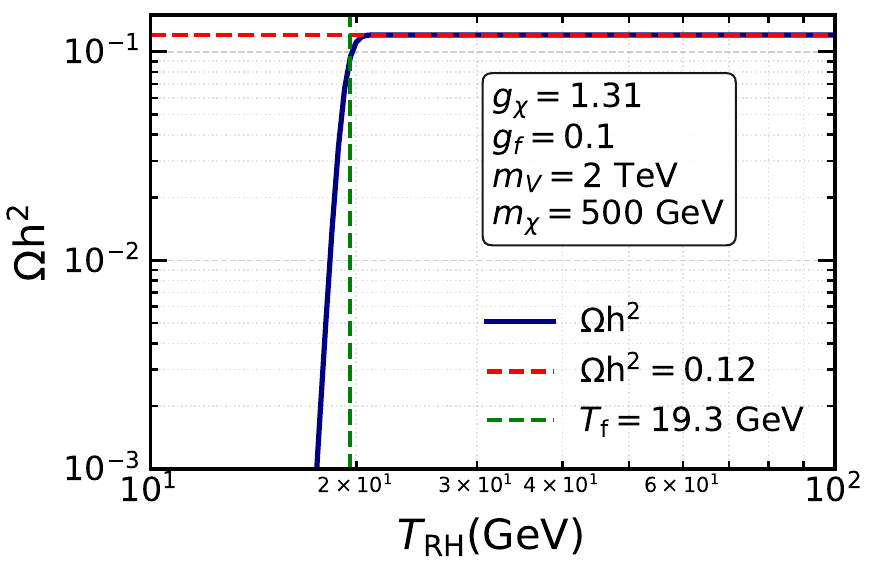}}
\caption{The freeze-out relic abundance in the finite-reheating
  scenario is shown for two representative benchmark points: (a)
  $m_V=500~{\rm GeV}$, $m_\chi=2~{\rm TeV}$, $g_f=0.1$, and (b)
  $m_V=2~{\rm TeV}$, $m_\chi=500~{\rm GeV}$, $g_f=0.1$. The observed relic abundance is obtained for $g_\chi=0.872$ and $g_\chi=1.31$ in cases (a) and (b), respectively. The
  effect of $T_{\rm RH}$ on the relic density becomes relevant only
  for $T_{\rm RH}\lesssim T_f$. Since $T_f\sim m_\chi/20$, the two
  benchmark points correspond to different freeze-out temperature
  scales, determined primarily by the DM mass. For the first
  benchmark, $T_f\sim70~{\rm GeV}$, while for the second benchmark,
  $T_f\sim20~{\rm GeV}$.}
\label{plot:fo}
\end{figure}

The resulting relic abundance as a function of $T_{\rm RH}$ is shown
in Fig.~\ref{plot:fo}. For both benchmarks, the relic abundance approaches the
standard radiation-dominated freeze-out prediction when $T_{\rm RH}$
is sufficiently above the freeze-out temperature. In this regime, the
interaction rate is larger than the
post-reheating Hubble expansion rate,
allowing the DM population to attain chemical equilibrium before
decoupling. Consequently, the subsequent evolution loses
virtually all memory of the initial reheating
temperature, and the conventional freeze-out result is recovered.


A qualitatively different behavior emerges once $T_{\rm RH}$
approaches the freeze-out scale. In this regime, the available
temperature interval between reheating and decoupling becomes too
short for the DM population to fully establish chemical
equilibrium. Lowering $T_{\rm RH}$ further therefore changes the
initial departure from equilibrium and leads to a progressively
smaller relic abundance. The onset of this deviation occurs at $T_{\rm
  RH}\sim T_f$, with $T_f\simeq73~{\rm GeV}$ for benchmark (a) and
$T_f\simeq20~{\rm GeV}$ for benchmark (b). The different values
primarily reflect the different DM masses, and numerical
  determinations are consistent with the usual expectation $T_f\sim
m_\chi/20$. The observed fall in the relic density in
Fig. \ref{plot:fo} is consistent with the analytical estimate
provided in Eq. \ref{eq:yield4}.

The freeze-out results, thus, provide a complementary perspective on the
role of a finite reheating temperature. Unlike freeze-in, for which
lowering $T_{\rm RH}$ progressively removes the high-temperature
contribution to DM production, the standard freeze-out abundance
remains essentially insensitive to $T_{\rm RH}$ as long as chemical
equilibrium can be established after reheating. A finite reheating
temperature becomes relevant only when it approaches the temperature
at which the DM would otherwise decouple, thereby preventing the
system from attaining the equilibrium state required for conventional
freeze-out.

\subsection{Dependence on the Coupling Strength}
\label{ssec:coup}

Having established the dependence of the relic abundance on the
reheating temperature in both the freeze-in and freeze-out regimes, we
now examine the complementary question of how the DM coupling required
to reproduce the observed relic abundance varies with $T_{\rm RH}$. As
before, we fix the coupling to the SM fermions at $g_f=0.1$ and
retain the two mass-benchmark scenarios considered above.

\begin{figure}[!ht]
\subfloat[]{\includegraphics[width = 2.8in]{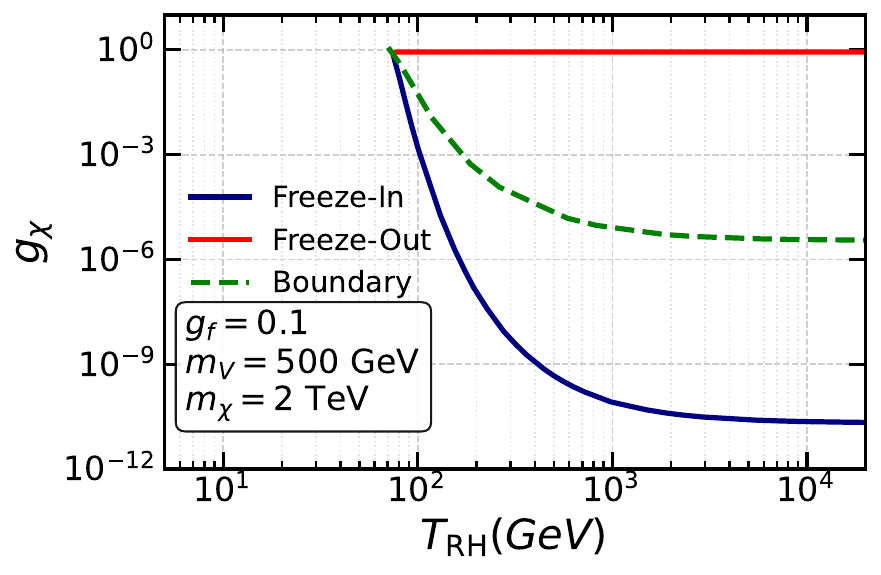}} \quad
\subfloat[]{\includegraphics[width = 2.8in]{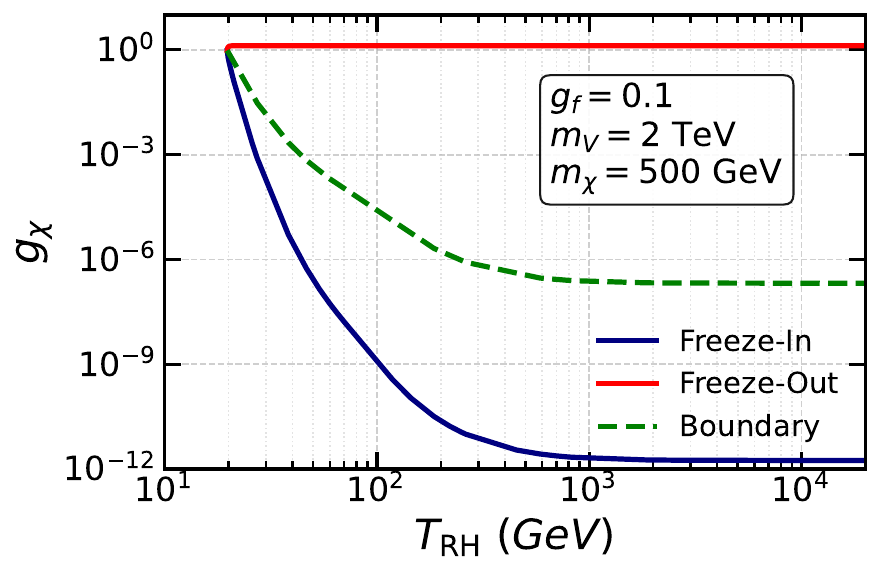}}
\caption{Plot of the coupling constant $g_\chi$ as a function of
  $T_{\rm RH}$ required to reproduce the observed relic abundance for
  two benchmark scenarios: (a) $m_V=500~{\rm GeV}$, $m_\chi=2~{\rm
    TeV}$, $g_f=0.1$, and (b) $m_V=2~{\rm TeV}$, $m_\chi=500~{\rm
    GeV}$, $g_f=0.1$. For the kinematically forbidden and allowed
  decay scenarios, respectively, the lowest possible $T_{\rm RH}$ at
  which the observed relic density can be achieved is different, with
  $T_{\rm RH}\sim60~{\rm GeV}$ for (a) and $T_{\rm RH}\sim20~{\rm
    GeV}$ for (b). These values coincide with the corresponding
  freeze-out temperatures for the two benchmarks, respectively. The
  green dashed line indicates the coupling at which the production
  mechanism switches from freeze-in to freeze-out as $g_\chi$
  increases.}
\label{plot:couplig}
\end{figure}

The resulting values of $g_\chi$ as a function of $T_{\rm RH}$ are
shown in Fig.~\ref{plot:couplig}. Two
distinct branches are apparent at sufficiently large reheating
temperatures. The red branch corresponds to the thermal freeze-out
solution, while the blue branch corresponds to the freeze-in
solution. As expected, in the freeze-out regime, the relic abundance
is insensitive to $T_{\rm RH}$ as long as it is sufficiently above the freeze-out
scale. Consequently, the magnitude of the coupling required to
reproduce the observed relic abundance remains essentially constant
and approaches its standard high-$T_{\rm RH}$ value. The freeze-in
branch exhibits the opposite behavior. Lowering $T_{\rm RH}$ removes
an increasingly large fraction of the thermal phase space available
for DM production, thereby suppressing the freeze-in yield. To
compensate for this suppression and maintain the observed relic
abundance, a progressively larger value of $g_\chi$ is required.

The green curve in Fig.~\ref{plot:couplig} delineates the transition between the non-equilibrium and thermal regimes. For fixed $m_\chi$, $m_V$, and $g_f$, it approximately corresponds to the coupling for which the DM interaction rate becomes comparable to the Hubble expansion rate,
\[
\Gamma_\chi(T)\equiv n_{\chi,\mathrm{eq}}(T)\langle\sigma v\rangle \sim H(T),
\]
and therefore marks the onset of efficient thermalization. The precise location of the transition is determined by the full Boltzmann evolution, since the DM need not thermalize instantaneously at the point where $\Gamma_\chi/H\simeq1$. Nevertheless, this criterion provides a useful physical interpretation of the boundary shown in the figure.

\begin{figure}[!ht]
\centering
\subfloat[]{\includegraphics[width = 3.8in]{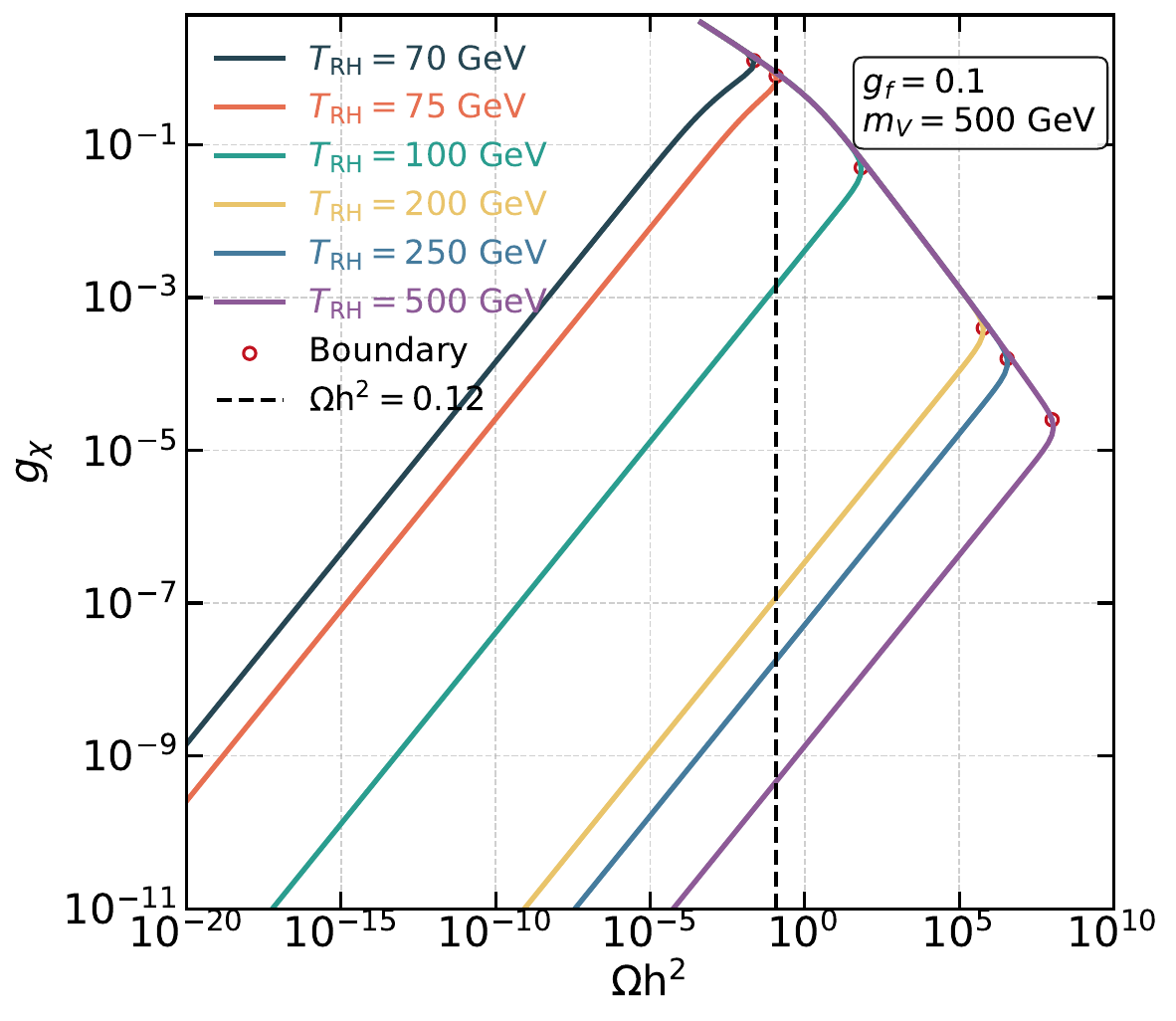}} 
\caption{Variation of the relic abundance, $\Omega_\chi h^2$, with the coupling $g_\chi$ for representative values of the reheating temperature, $T_{\rm RH}$. The red circles along each curve indicate the transition between the freeze-in and freeze-out regimes.}
\label{plot:couplig_2}
\end{figure}

Below (or to the left of) the blue line, the coupling $g_\chi$
  is too small to drive freeze-in efficiently enough and the relic
  density is too small. As we increase the coupling keeping $T_{\rm
    RH}$ constant, progressively larger fractions of the progenitors
  go into the DM, thereby increasing its abundance. The latter reaches
  a maximum along the green (dashed line). For still larger $g_\chi$,
  the pair annihilation of the DM is progressively rendered more
  important, leading to a fall in the relic density. In other words,
  one has entered the regime wherein freeze-out starts to
  dominate. Above the red line, the pair annihilation is too efficient
  for sufficient abundance to survive.

Equivalently, as $T_{\rm RH}$ is lowered, the freeze-in solution
moves towards larger values of $g_\chi$ and eventually reaches this
thermalization boundary. At this point, the coupling required to
obtain the observed relic abundance coincides with the largest relic
abundance attainable at the boundary between the freeze-in and
freeze-out regimes. Further lowering $T_{\rm RH}$ leads to a
qualitatively different situation. The relic abundance evaluated along
the thermalization boundary falls below the observed value, leaving no
coupling for which the correct relic density can be
reproduced. Indeed, reducing $g_\chi$ below the boundary moves the
system into the freeze-in regime, where the already suppressed
production rate yields an insufficient abundance. Increasing $g_\chi$
above the boundary instead drives the system into the freeze-out
regime, where the enhanced annihilation rate lowers the final
abundance below the observed value. The observed relic density
therefore cannot be recovered from either side of the transition and
the two mechanisms cease to provide viable solutions once the
reheating temperature becomes sufficiently low.

This behavior is clearly illustrated in Fig.~\ref{plot:couplig_2}, which shows the relic abundance as a function of $g_\chi$ for six representative values of $T_{\rm RH}$. The red circles indicate the transition between the freeze-in and freeze-out regimes along each curve. For a fixed reheating temperature, the relic abundance initially increases with $g_\chi$ in the freeze-in regime, reaches a maximum around the transition, and subsequently decreases as the system enters the freeze-out regime. As $T_{\rm RH}$ is lowered, this maximum attainable relic abundance decreases. Eventually, at a critical reheating temperature, the maximum abundance becomes equal to the observed relic abundance, causing the freeze-in and freeze-out solutions to merge into a single solution. Below this temperature, the maximum relic abundance attainable for any value of $g_\chi$ falls below the observed value, and hence no viable relic-density solution exists. This establishes a minimum reheating temperature, $T_{\rm RH}^{\rm min}$, for a given choice of $m_\chi$, $m_V$, and $g_f$, below which the observed DM relic abundance cannot be reproduced for any value of $g_\chi$. The resulting lower bound on $T_{\rm RH}$ is one of the central phenomenological consequences of the finite reheating temperature scenario considered in this work.



\section{Summary and conclusions}
\label{sec:sum}

In recent years, a low-reheating temperature has often been proffered
as a panacea for the ills of DM models. Given this, we reinvestigate
the cosmological evolution of the dark-matter density when the thermal
bath is established at a finite temperature $T_{\rm RH}$ (as opposed
to a very high one), while retaining an otherwise standard
radiation-dominated cosmological history. $T_{\rm RH}$, then, sets the
maximum temperature available for thermal production and hence
determines which all stages of the production history are accessible.

While the qualitative features of our conclusions are largely
model-independent, we illustrate our arguments in terms of a very
simplified model wherein a Dirac field $\chi$ constitutes the DM with
its interactions with the SM particles mediated by a single gauge
field $V$. This gives rise to a particularly transparent sequence in
the freeze-in scenario, as $T_{\rm RH}$ is reduced holding the
particle physics parameters unchanged. When $V \to \bar \chi \chi$ is
kinematically accessible, expectedly, it constitutes the dominant
contribution to the relic density. As $T_{\rm RH}$ is lowered, $V$
production itself is suppressed, thereby dampening this
contribution. In this regime, SM-pair annihilation to $\bar \chi \chi$
via an off-shell $V$ dominates. As $T_{\rm RH}$ is lowered further,
this cross section too falls owing to the inherent $s$-channel
suppression. The extent to which this intermediate regime is realized
is dictated by the separation between the mediator and DM mass
scales. Finally, for low enough $T_{\rm RH}$, the available energy is
too small to allow for efficient DM production and the relic abundance
falls off exponentially.

It is instructive to consider both the freeze-in and freeze-out
mechanisms for the same particle physics model, albeit with vastly
differing coupling constants for the two cases.  At high $T_{\rm RH}$,
the same particle-physics model can admit both a feeble-coupling
solution in which the observed abundance is accumulated without
thermalization and a strong-coupling solution in which the abundance
is set by thermal freeze-out. Lowering $T_{\rm RH}$ progressively
depletes the maximum abundance attainable through freeze-in, while
eventually also preventing the system from reaching the equilibrium
trajectory required for freeze-out. The two solutions can, consequently,
converge and disappear below a certain $T_{\rm RH}^{\rm min}$. In
other words, a sufficiently low reheating temperature does not merely
shift the preferred coupling: it can remove the relic-density solution
altogether. This provides a useful way of viewing low-reheating
cosmologies--as a restriction on the accessible production
history--and suggests that, even in an otherwise standard
post-reheating Universe, the requirement of obtaining the observed DM
abundance can translate into a lower bound on the reheating
temperature for a specified dark-sector realization.

More broadly, the analysis illustrates how dark-matter phenomenology
can provide a probe of the otherwise poorly constrained thermal
history of the Universe prior to BBN. If a dark-sector realization
with independently established masses and couplings were identified,
the requirement of reproducing the observed relic abundance could be
used to infer a lower limit on $T_{\rm RH}$ within the assumed
instantaneous-reheating cosmology. In this sense, information about
the particle nature of dark matter can be translated into information
about the temperature at which the thermal bath was established,
providing a possible connection between particle-physics observables
and the pre-BBN cosmological history.

\acknowledgments DC acknowledges the IoE, University of Delhi grant
IoE/2025-26/12/FRP. The work of RS during the initial stage of this project was supported by the Anusandhan National Research Foundation (ANRF) under Grant No. CRG/2023/008234. During the final stage of the project, the work of RS was fully supported by the Anusandhan National Research Foundation, Advanced Research Grant (ANRF/ARG/2025/005801/PS).

\newpage
\bibliography{references}
\end{document}